\documentstyle[11pt]{article}
\title{ Inelastic Final-State Interactions of $B \rightarrow
        VV \rightarrow \pi K$ Processes 
        \thanks{Supported in part by National Natural Science 
	        Foundation of China } }
\author{ Dong-Sheng Du$\rm ^{a,b}$,
         Xue-Qian Li$ \rm ^{a,c}$, 
         Zheng-Tao Wei$ \rm ^{b}$, 
         Bing-Song Zou$ \rm ^{d}$
         \thanks{E-mail address: duds@bepc3.ihep.ac.cn;
	 weizt@hptc5.ihep.ac.cn.}\\ 
         $\rm ^a$ \it{CCAST(World Laboratory),
         ~~P.O.Box $8730$, Beijing $100080$, China}\\
         $\rm ^b$ \it{Institute of High Energy Physics, 
         P.O.Box $918(4)$, Beijing $100039$, China}\\
         $\rm ^c$ \it{Department of Physics, Nankai University, 
         Tianjin $300071$, China} \\
         $\rm ^d$ \it{Queen Mary and Westfield College, London 
         E1 $4$NS, UK} }

\date{}
\begin{document}
\newcommand{\Adir}{\mbox{${\cal A}^{\rm dir}_{\rm CP}$}}

\maketitle
\vspace*{0.3cm}


\section*{Abstract}
We study the final-state interactions in $B \rightarrow \pi K$ decays
through $B \rightarrow VV \rightarrow \pi K$ processes where
the inelastic rescattering occurs via single pion exchange. 
The next-to-leading order low energy effective  Hamiltonian and BSW 
model are used to evaluate the weak transition matrix elements 
and the final-state interactions. We found that the final-state 
interaction effects in $B \rightarrow \rho K^* \rightarrow \pi K$ 
processes are significant. The Fleischer-Mannel relation about 
the CKM angle $\gamma$ may be significantly modified.

\newpage
\section*{1. Introduction}

Final-state interactions (FSI) play a great role in many physical 
processes, especially in the weak decays which are the one of the 
focus of recent interests. Their  effects lie in two aspects. 
First, strong phases in weak decay 
amplitudes are generated by the final-state interactions, they may  
contribute the strong phases for the direct CP asymmetries; 
second, FSI effects may significantly change the theoretical
predications for certain quantities. Study on final-state
interactions would definitely need information about the
non-perturbative effects of low energy 
hadron interactions. Unless we can correctly evaluate the FSI effects, 
it is impossible to extract reliable information about the reaction 
mechanism or new physics from the data. Understanding the final-state 
interaction effects in weak decays is not only crucially important, 
but also  is a challenging and difficult task in both theory and 
phenomenology. Up to now, definite quantitative analysis has not 
been accessible yet. The estimate of the final-state 
interactions is centered on some particular cases  where the 
symmetry relations can be applied, or one can use
some simplified medels for example, the Regge pole
or single-pion exchange etc. to do the job.

The origin of CP violation in the Standard Model comes from the
complex phase of the Cabibbio-Kobayashi-Maskawa (CKM) matrix.
In general, for three generation quark families the CKM matrix elements
form a unitarity triangle.
So, reasonable extraction of each angles ($\alpha, \beta, 
\gamma$) is extremly important for testing the Standard Model.
Among the three angles, to extract $\gamma$ is  most difficult. 
Some methods \cite{GRL} have been put 
forward for this purpose.  But all the methods are either too
complicated or impractical from experimental point of view. 
Last year, the CLEO collaboration  reported the combined ratios
for $ B \rightarrow \pi K $ decays \cite{CLEO}:
$$
\begin{array}{ll}
BR(B^{\pm}\rightarrow \pi^{\pm}K) &=
  (2.3^{+1.1+0.2}_{-1.0-0.2}\pm 0.2)\times 10^{-5}\\
BR(B_d\rightarrow \pi^{\mp} K^{\pm}) &=
  (1.5^{+0.5+0.1}_{-0.4-0.1}\pm 0.1)\times 10^{-5} 
\end{array}
$$
Fleischer and Mannel \cite{Fleischer} give a bound relation 
on the CKM  angle $\gamma$ based on the above results: 
$R\geq sin^2\gamma$, 
where $R=\frac{BR(B_d \rightarrow \pi^{\mp} K^{\pm})}{BR(B^{\pm}
\rightarrow \pi^{\pm)} K}=0.65\pm 0.40$, $\gamma\equiv Arg(V_{ub}^*)$.
In their work, final-state interactions were neglected. 
However, the final-state interactions in such channels may be
important and cannot be ignored.  Some authors \cite{Delepine}
have studied the final-state interaction effects on
$B\rightarrow \pi K$ decays. Their studies are based on the Regge 
pole theory. In Ref.\cite{Zou}, the authors study the 
final-state interactions due to single pion exchange in 
$D\rightarrow VP$ processes. Their results show that the single 
pion exchange can be significant. It is believed that even though 
the two schemes describe the process based on different physical 
considerations, in practice, each of them works well, just because
the uncertaities are partly compensated by proper
selection of some phenomenological parameters in these schemes.

In this paper, we study the final-state interactions in 
$B\rightarrow VV\rightarrow \pi K$ due to single pion exchange. 
In Standard Model calculation, the decay of $B\rightarrow \rho K^*$ 
has the same order amplitude as that of $B \rightarrow \pi K$.
In fact, both $K^*\rightarrow  K \pi$ and $\rho \rightarrow \pi\pi$ 
are the dominant strong decay  channels, so that the single pion
exchange mechanism may dominate the final state interactions.
In our processes,
the exchanged pion is in the t-channel.
There are other two-vector-meson states which can rescatter into the
$\pi K$ final state by  pion exchange. 
But these intermediate mesons have smaller couplings to $\pi K$ 
and $\pi \pi$ than $K^* \rightarrow \pi K$ and $\rho \rightarrow 
\pi\pi$; also they are heavier than $\rho$ and $K^*$ and will result 
in larger t-value which will further reduce their contribution.
So, the final-state interactions in $B \rightarrow \rho K^* 
\rightarrow \pi K$ processes may be the largest in the processes of
$B\rightarrow VV \rightarrow \pi K$.  
We use the method which was presented in Ref.\cite{Zou}. 
The next-to-leading order low energy effective  
Hamiltonian and BSW model are used to evaluate the weak 
transition matrix elements and the final-state interactions. 
We find that the final-state interaction effects in
$B\rightarrow \rho K^* \rightarrow \pi K$ are significant. 

\section*{2. The formulation for FSI in effective Hamiltonian}

To evaluate the decays of $B\rightarrow \pi K$, we use the 
next-to-leading order low energy effective Hamiltonian and 
BSW model. The next-to-leading order low energy 
effective Hamiltonian describing $|\Delta B|=1$ transitions
is given at the renormalization scale $\mu=O(m_b)$ as \cite{Buras}:  
$$
{\cal H}_{eff}(|\Delta B|=1) = \frac{G_F}{\sqrt{2}}
\left[\sum_{q=u,c}v_q \left\{ Q_1^qC_1(\mu)+Q_2^qC_2(\mu)
                       +\sum_{k=3}^{10} Q_k C_k(\mu)\right\}\right]+H.C.
\eqno(1)
$$
The CKM factors $v_q$ are defined as $v_q=V_{qs}^*V_{qb}$, where
$q=u, ~ c$. 

The ten operators $Q_1^u$, $Q_2^u$, $Q_3,\ldots,Q_{10}$ are known as the
following forms:
$$
\begin{array}{ll}
Q_1^u=  (\bar{q}_{\alpha}u_{\beta})_{V-A}(\bar{u}_{\beta}b_{\alpha})_{V-A} &
Q_2^u=  (\bar{q}u)_{V-A}(\bar{u}b)_{V-A}\\
Q_{3(5)}=  (\bar{q}b)_{V-A}\displaystyle\sum_{q'}(\bar{q}'q')_{V-A(V+A)} &
Q_{4(6)}=  (\bar{q}_{\alpha}b_{\beta})_{V-A}
      \displaystyle\sum_{q'}(\bar{q}'_{\beta}q'_{\alpha})_{V-A(V+A)} \\
Q_{7(9)}=  \frac{3}{2}(\bar{q}b)_{V-A}
      \displaystyle\sum_{q'}e_{q'}(\bar{q}'q')_{V+A(V-A)} &
Q_{8(10)}=  \frac{3}{2}(\bar{q}_{\alpha}b_{\beta})_{V-A}
      \displaystyle\sum_{q'}e_{q'}(\bar{q}'_{\beta}q'_{\alpha})_{V+A(V-A)} \\
\end{array}
\eqno(2)
$$
where $Q_1^u$ and $Q_2^u$ are the current-current operators, and
the current-current operators $Q_1^c$ and $Q_2^c$ can be obtained from
$Q_1^u$ and $Q_2^u$ through the substitution of $u\rightarrow c$.
$Q_3,\ldots,Q_6$ are the QCD penguin operators, whereas $Q_7,\ldots,Q_{10}$
are the electroweak penguin operators. The quark $q=s$  for
$b \rightarrow s$ transition;   
$(V \pm A)$ refer to $\gamma_{\mu}(1\pm \gamma_5)$.  

The matrix elements are: 
$$
<{\bf Q}^T(\mu)\cdot {\bf C}(\mu)>
\equiv<{\bf Q}^T>_0\cdot{\bf C'}(\mu)
\eqno(3)
$$
where $<{\bf Q}>_0$ denote
the tree level matrix elements of these operators,
and  ${\bf C'}(\mu)$ are defined as
$$
\begin{array}{llll}
C'_1=~\overline{C}_1, &C'_2~=~\overline{C}_2, &
C'_3=~\overline{C}_3-P_s/3, &C'_4~=~\overline{C}_4+P_s,\\
C'_5=~\overline{C}_5-P_s/3, & C'_6~=~\overline{C}_6+P_s,&
C'_7=~\overline{C}_7+P_e,   &C'_8~=~\overline{C}_8, \\
C'_9=~\overline{C}_9+P_e,   &C'_{10}~=~\overline{C}_{10},
\end{array} \eqno(4)
$$
where $P_{s,e}$ are given by
$$
\begin{array}{rl}
P_s&=~\frac{\alpha_s}{8\pi}\overline{C}_2(\mu)\left[\frac{10}{9}-
       G(m_q,q,\mu)\right],\\
P_e&=~\frac{\alpha_{em}}{9\pi}\left(3\overline{C}_1+\overline{C}_2(\mu)
      \right)\left[\frac{10}{9}-G(m_q,q,\mu)\right],\\
G(m,q,\mu)&=~-4\int_0^1 dx~x(1-x)ln\displaystyle\left[\frac{m^2-x(1-x)q^2}
{\mu^2}\right],
\end{array} \eqno(5)
$$
here $q$ represents  $u,~c$, and $q^2=m_b^2/2$.
The numerical values of the renormalization scheme
independent Wilson Coefficients $\overline{C}_i(\mu)$ at $\mu=O(m_b)$
are \cite{Hxg}
$$
\begin{array}{llll}
\bar{ c}_1=-0.313, & \bar {c}_2=1.150, &\bar{ c}_3=0.017, &
\bar{ c}_4=-0.037, \\
\bar{ c}_5=~0.010,  & \bar{ c}_6=-0.046, &
\bar{ c}_7=-0.001\cdot \alpha_{em}, \\
\bar{ c}_8=0.049\cdot\alpha_{em}, &
\bar{ c}_9=-1.321\cdot\alpha_{em}, &
\bar{ c}_{10}=0.267\cdot\alpha_{em}.
\end{array}\eqno(6)
$$

\vspace{0.3cm}
\begin{large}
\noindent
(1) Without the final-state interactions 
\end{large}

\vspace{0.3cm}
\noindent
In $B^-_u \rightarrow \pi^- \bar{K^0} $ decay, only penguin diagrams
contribute. As commonly agreed in the present literatures, the
annihilation  diagram contributions are
neglected because of $V_{ub}$ and the form factor suppression.
In $\bar{B^0_d} (b\bar d) \rightarrow \pi^+ K^- $ decay,
both tree and penguin
diagrams contribute. The amplitudes of these two decays are:
$$
\begin{array}{ll}
A^{dir}(B^-_u \rightarrow \pi^- \bar{K^0}) 
 &=\frac{G_F}{\sqrt{2}}\sum\limits_{q=u,c}v_q 
 [a_3+\frac{2M_{K^0}^2}{(m_s+m_d)(m_b-m_d)}(a_5-\frac{1}{2}a_7)
 -\frac{1}{2}a_9 ]M^{\bar{K^0}\pi^-}\\
A^{dir}(\bar{B^0} \rightarrow \pi^+ K^-) 
 &=\frac{G_F}{\sqrt{2}}\sum\limits_{q=u,c}v_q [a_1\delta_{uq}+a_3
 +\frac{2M_{K^-}^2}{(m_s+m_u)(m_b-m_u)}(a_5+a_7)+a_9 ]M^{K^+\pi^-}
\end{array}
\eqno(7)
$$
where ``$dir$'' means direct decay without final-state interactions, 
and
$$
\begin{array}{l}
M^{\bar{K^0}\pi^-} 
  \equiv <\bar{K^0}|(\bar{s}d)_{V-A}|0><\pi^-|(\bar{d}b)_{V-A}|B_u^->
  =-i f_K F_0^{B^-_u\pi}(M_{K^0}^2)(M_{B_u}^2-M_{\pi^-}^2)\\
M^{K^-\pi^+}
  \equiv <K^-|(\bar{s}u)_{V-A}|0><\pi^+|(\bar{u}b)_{V-A}|\bar{B^0}>
  =-i f_K F_0^{B^0\pi}(M_{K^-}^2)(M_{B^0}^2-M_{\pi^+}^2)
\end{array}
\eqno(8)
$$
and
$$
\begin{array}{ll}
a_{2i-1}&=C_{2i-1}'/N_c+C_{2i}' \\
a_{2i}  &=C_{2i-1}'+C_{2i}'/Nc
\end{array}
\eqno(9)
$$

Becsuse $|v_c|/|v_u|>>1$, the tree diagram contributions to the decay
amplitude are small compared to the Penguin diagrams. The factorization
approximation and BSW model \cite{BSW} are used to evaluate the matrix
elements in Eq.(8). Table.1 gives  the calculation results in the
standard method for the above two direct decays. The non-factorization
effects have been considered by the choice of $N_c=\infty,
3, 2$. 

\vspace{0.5cm}
\begin{large}
\noindent
(2) With the final-state interactions 
\end{large}

\vspace{0.3cm}
\noindent
In $B\rightarrow \rho K^*\rightarrow \pi K$ processes, the exchanged 
pions can be neutral and charged as shown in Fig.1.   
For charged pion exchange the decay $B_u^{\mp} 
\rightarrow \pi^{\mp}K$ can get the tree diagram contribution 
in the intermidiate state $B\to \rho K^*\to \pi K$. 

We take the decay $B_u^- \rightarrow \rho^- \bar{K^{*0}} \rightarrow 
\pi^- \bar{K^0}$ as an example to show how to calculate the
final-state interaction effects. 

The amplitude for $B_u^- \rightarrow \rho^- \bar{K^{*0}}$ decay is:
$$
A(B^-_u \rightarrow \rho^- \bar{K^{*0}}) 
 =\frac{G_F}{\sqrt{2}}\sum\limits_{q=u,c}v_q 
 (a_3-\frac{1}{2}a_9 )M^{\bar{K^{*0}}\rho^-}
\eqno(10)
$$
where
$$
\begin{array}{ll}
M^{\bar{K^{*0}}\rho^-} &\equiv <\bar{K^{*0}}|(\bar{s}d)_{V-A}|0>
  <\rho^-|(\bar{d}b)_{V-A}|B_u^->\\
 &=\frac{2M_{K^{*0}}}{M_{B_u}+M_{\rho^-}}f_{K^*}V^{B_u\rho}(M_{K^{*0}}^2)
   \epsilon_{\mu\nu\rho\sigma}\epsilon_{K^{*0}}^{\mu}
   \epsilon_{\rho^-}^{\nu}p_{K^{*0}}^{\rho}p_{\rho^-}^{\sigma}\\
 &~~+iM_{K^{*0}}(M_{B_u}+M_{\rho^-})f_{K^*}A_1^{B_u\rho}(M_{K^{*0}}^2)
   (\epsilon_{K^{*0}}\cdot \epsilon_{\rho^-})\\
 &~~-i\frac{2M_{K^{*0}}}{M_{B_u}+M_{\rho^-}}f_{K^*}A_2^{B_u\rho}(M_{K^{*0}}^2)
   (\epsilon_{K^{*0}}\cdot p_B) (\epsilon_{\rho^-}\cdot p_B)
\end{array}
\eqno(11)
$$

As in Ref.\cite{Zou}, the single pion exchange in the t-channel makes a
significant contribution to the FSI and would  dominate.
To get the absoptive part of the loop as shown in Fig.1a, 
the way to make cuts is: let the $\rho^-$ and $\bar{K}^{0*}$ be
on-shell, and leave the exchanged pion to be off-shell.

In the center of mass frame of $B_u^-$ where
$p_{_{B_u}}=(M_{B_u},0)$, the matrix
element in Eq.(11) is recast into the following form for the process
$B_u^- \rightarrow \rho^- \bar{K^{*0}} \rightarrow 
\pi^- \bar{K^0}$
$$
\begin{array}{ll}
M_{FSI}^{\bar{K^0}\pi^-} &=\frac{1}{2}\int \frac{d^3 p_1}{(2\pi)^3 2E_1}
   \frac{d^3 p_2}{(2\pi)^3 2E_2}(2\pi)^4 \delta^4(p_1+p_2-p_B)
   <\pi^-\bar{K^0}|S|\rho^-\bar{K^{*0}}>M^{\bar{K^{*0}}\rho^-}\\
 &=\int \frac{|\stackrel{\rightarrow}{p}|}{16\pi M_{B_u}}d(cos\theta)
  \frac{iF(p_{\pi^0}^2)}{(p_{\pi^0}^2-M_{\pi^0}^2)}
  [iM_{K^{*0}}(M_{B_u}+M_{\rho^-})f_{K^*}A_1^{B_u\rho}(M_{K^{*0}}^2)
  \cdot H_1 \\
 &~~-i\frac{2M_{K^{*0}}}{M_{B_u}+M_{\rho^-}}M_{B_u}^2
  f_{K^*}A_2^{B_u\rho}(M_{K^{*0}}^2)\cdot H_2]
\end{array}
\eqno(12)
$$
where S is the S-matrix of strong interaction, 
$\theta$ is the angle between $\stackrel{\rightarrow}{p_1} $and  
$\stackrel{\rightarrow}{p_3}$, and 
$$
\begin{array}{ll}
H_1 &=-4g_{\rho\pi\pi}g_{K^*K\pi}[(p_3\cdot p_4)
    -\frac{(p_2\cdot p_3)(p_2\cdot p_4)}{M_2^2 }\\
   &~~~-\frac{(p_1\cdot p_3)(p_1\cdot p_4)}{M_1^2 }
    +\frac{(p_1\cdot p_2)(p_2\cdot p_3)(p_2\cdot p_4)}{M_1^2 M_2^2}]\\
H_2 &=-4g_{\rho\pi\pi}g_{K^*K\pi}[(p_3^0 p_4^0)
    -\frac{(p_2^0 p_3^0)(p_2\cdot p_4)}{M_2^2 }\\
   &~~~-\frac{(p_1^0 p_4^0)(p_1\cdot p_3)}{M_1^2 }
    +\frac{(p_1^0 p_2^0)(p_2\cdot p_3)(p_2\cdot p_4)}{M_1^2 M_2^2}]\\
\end{array}
\eqno(13)
$$
We set $p_1=p_{_{K^*}}, p_2=p_{\rho}, p_3=p_{_K}, p_4=p_{\pi},
M_1=M_{K^*}, M_2=M_{\rho}, M_3=M_K, M_4=M_{\pi}$.

So, the amplitude of $B_u^- \rightarrow \rho^- \bar{K^{*0}} \rightarrow 
\pi^- \bar{K^0}$ is: 
$$
A^{FSI}(B^-_u \rightarrow \pi^- \bar{K^{0}}) 
 =\frac{G_F}{\sqrt{2}}\sum\limits_{q=u,c}v_q 
 (a_3-\frac{1}{2}a_9 )M_{FSI}^{\bar{K^{0}}\pi^-}
\eqno(14)
$$

The factor $F(p_{\pi^0}^2)$ in Eq.(12) is an off-shell form factor for the 
vertices $K^*K\pi$ and $\rho \pi \pi$. We take 
$F(p_{\pi^0}^2)=(\frac{\Lambda^2-m_{\pi^0}^2}
{\Lambda^2-p_{\pi^0}^2})$ as in Ref.\cite{Zou}, 
where $\Lambda=1.2-2.0GeV$.  

\section*{3. The numerical results}

The parameters such as meson decay constants, form factors and 
quark masses needed in our calculations are taken as:

meson decay constant \cite{PDG} \cite{Guo}:

$
f_{\pi}=0.13GeV,~ 
f_K=0.16GeV, ~
f_{\rho^0}^{u\bar{u}}=0.156GeV, ~
f_{K^*}=0.221GeV;
$

form factor \cite{Guo}:
$$
\begin{array}{llll}
F_0^{B\pi}(0)=0.333,&
F_1^{B\pi}(0)=0.333, &
V^{BK^*}(0)=0.369, &
A_1^{BK^*}(0)=0.328,\\
A_2^{BK^*}(0)=0.331, &
V^{B\rho}(0)=0.329, &
A_1^{B\rho}(0)=0.283,& 
A_2^{B\rho}(0)=0.283; 
\end{array}
$$
 
effective strong coupling constants \cite{Zou}:
$g_{K^* K\pi}=5.8$,~~$g_{\rho \pi \pi}=6.1$; 

$\Lambda$ in the off-shell form factor $F(p_{\pi^0}^2)$:
$\Lambda=1.5GeV$; 

quark mass \cite{PDG}: 
 
$m_u=0.005GeV$,
$m_d=0.01GeV$,
$m_s=0.2GeV$,
$m_c=1.5GeV$,
$m_b=4.5GeV$;

the Wolfenstein CKM parameters \cite{Ali}:

$\lambda=0.22$,~
$A=0.8$,~
$\eta=0.34$,~
$\rho=-0.12$.

Due to the non-factorization effects,
it is hard to choose the value of $N_c$, so
all three cases are taken into account: $N_c=\infty, 3, 2$. 

The corresponding numerical results are presented in Table 1 and Table 2.

\section*{4. The Constraints on $\gamma$ and ${\cal{A}}^{dir}_{cp}$ }

(i) Without final-state intearaction

For direct decays of $B\to \pi K $, the ampiltudes are:
$$
\begin{array}{ll}
A^{dir}(B^+ \rightarrow \pi^+ K^0)=
A^{+}_{cs}-A^{+}_{us}e^{i\gamma}e^{i\delta_+},  &\\
A^{dir}(B^-\rightarrow \pi^- \bar{K^0})=
A^{+}_{cs}-A^{+}_{us}e^{-i\gamma}e^{i\delta_+}, &\\
A^{dir}(B^0\rightarrow \pi^- K^+)=
A^{0}_{cs}-A^{0}_{us}e^{i\gamma}e^{i\delta_0},  &\\
A^{dir}(\bar{B^0}\rightarrow \pi^+ K^-)=
A^{0}_{cs}-A^{0}_{us}e^{-i\gamma}e^{i\delta_0}  &\\
\end{array}
\eqno(15)
$$
where $\delta_0$ and $\delta_+$ are CP-conserving strong phases. 

The ratio $R$ is defined by: 
$$
\begin{array}{ll}
R &\equiv ~~ \frac{BR(B^0\rightarrow \pi^- K^+)+
  BR(\bar{B^0}\rightarrow \pi^+ K^-)}
  {BR(B^+ \rightarrow \pi^+ K^0)+
   BR(B^-\rightarrow \pi^- \bar{K^0})}\\
 &=~~(\frac{A_{cs}^0}{A_{cs}^+})^2 \frac
   {1-2r_0\cos\gamma\cos\delta_0 +r_0^2 } 
   {1-2r_+\cos\gamma\cos\delta_+ +r_+^2}
\end{array}
\eqno(16)
$$
where $r_0=A_{us}^0/A_{cs}^0$, $r_+=A_{us}^+/A_{cs}^+$. 
When neglecting the electro-weak penguin diagram contributions,
$A_{us}^+=0$, i.e. $r_+=0$. According to SU(2) isospin symmetry 
of strong interaction. $A_{cs}^0/A_{cs}^+\approx 0$. So, 
$$
R=1-2r_0 \cos\gamma \cos\delta+r_0^2. 
\eqno(17)
$$
Following Fleischer and Mannel[3], we can obtain the
inequality
$$
R\geq \sin^2 \gamma
\eqno(18)
$$
This is the Fleischer-Mannel relation. 

Direct CP violation in $B^{\pm}\to \pi^{\pm}K$ is defined through
the CP asymmetry 
$$
\Adir \equiv 
\frac{BR(B^+\to\pi^+K^0)-BR(B^-\to\pi^-\bar{K^0})}
{BR(B^+\to\pi^+K^0)+BR(B^-\to\pi^-\bar{K^0})}
=\frac{2r_+\sin\gamma\sin\delta_+}
{1-2r_+\cos\gamma\cos\delta_++r_+^2}.
\eqno(19)
$$
where this "dir" means direct CP asymmetry. In direct decays, 
it is small from Eq.(19),
$$
\Adir \leq{\cal O}(\lambda^2).
\eqno(20)
$$

(ii) With the final-state interactions 

When considering the final-state interactions, 
the amplitude of $B\to \pi K$ is changed  to: 
$$
\begin{array}{ll}
~~~A(B^+\to \pi^+ K^0)\\
 =A^{dir}(B^+\to \pi^+ K^0)+
 A^{FSI}(B^+\to \rho^+ K^{*0}\to \pi^+ K^0)+ 
 A^{FSI}(B^+\to \rho^0 K^{*+}\to \pi^+ K^0)\\
=(A_{cs}^+ -A_{us}^+e^{i\gamma}e^{i\delta_+})
 (1+A_1e^{i\delta_1}+A_3e^{i\gamma}e^{i\delta_3}) \\
~~~A(B^0\to \pi^- K^+)\\
=A^{dir}(B^0\to \pi^- K^+)+
 A^{FSI}(B^0\to \rho^- K^{*+}\to \pi^- K^+)+ 
 A^{FSI}(B^0\to \rho^0 K^{*0}\to \pi^- K^+)\\
=(A_{cs}^0 -A_{us}^0e^{i\gamma}e^{i\delta_0})
 (1+A_2e^{i\delta_1}) 
\end{array}
\eqno(21)
$$
where $A_1$, $A_2$, $A_3$ are final-state interaction amplitudes, 
and $\delta_1$, $\delta_2$, $\delta_3$, are the strong phases
caused by the final-state interactions, i.e. the phase shifts of the inelastic
rescattering. The term  $A_3e^{i\gamma}e^{i\delta_3}$ is the contribution
from the tree diagram of $B^+\to \rho^0 K^{*+}\to \pi^+ K^0$.  
Our numerical calculations give $A_1=0.5$, $A_2=0.15$,
$A_3=0.05$, $\delta_1\approx 90^0$, $\delta_2\approx 90^0$. 
The strong phases are $90^0$, because we calculate only
the absorptive part of the hadron loop caused by the final-state interactions.
We will come back to this point later in the last section.
$A_1=0.5\geq A_2=0.15$ is encouraging for our mechanism, because
it can explain that the experimental branching ratio of 
$B^\pm \to \pi^{\pm}K$ is largger than that of
$B^0 \to \pi^{\mp}K^{\pm}$.

The ratio $R$ in Eq.(16) is changed into
$$
R=\frac{1+A_2^2+2A_2\cos\delta_2}{1+A_1^2+2A_1\cos\delta_1}
  (1-2r_0\cos\gamma\cos\delta_0+r_0^2)
\eqno(22)
$$
If we define $R'$  
$$
R'=1-2r_0\cos\gamma\cos\delta_0+r_0^2
\eqno(23)
$$
then from Eq.(22), (23),  
$$
R'=\frac{1+A_1^2+2A_1\cos\delta_1}{1+A_2^2+2A_2\cos\delta_2}R
\eqno(24)
$$
So, the Fleischer-Mannel relation is changed to 
$$
R'\geq \sin^2 \gamma
\eqno(25)
$$
When $A_1=0.5$, $A_2=0.15$, $\delta_1= 90^0$, $\delta_2= 90^0$, 
$R'=1.25R$. The bound relation in Eq.(18) is modified as much 
as $25\%$, when only the absorptive part of the hadron loop is under
consideration. No doubt, the dispersive part of the loop will also contribute
to the modification. Let us take an  example, when  the dispersive part of
the amplitude is equal to the absorptive part, $A_1=0.5\sqrt{2}$,
$A_2=0.15\sqrt{2}$, $\delta_1=45^0$, $\delta_2=45^0$, 
then $R'=1.58R $. So,  the bound relation is modified significantly 
as much as $58\%$.
      
For direct CP asymmetry, 
$$
\Adir\approx 2A_3\sin\gamma\sin\delta_+
\eqno(26)
$$ 
The final-state interaction can provide about $5\%-10\%$ direct CP  
asymmetry. But since the relative sign cannot be fixed by the theory,
we are unable to determine whether the correction is constructive or 
destructive.

\section*{5. Conclusion and discussion}

From the numerical results shown in Table 1 and Table 2, one can notice 
that the
final-state interactions due to the single pion exchange in $B\rightarrow 
\rho K^* \rightarrow \pi K$ processes are  $10\% - 30\%$ relative to the 
direct decay amplitude. This result is based on considering only the 
absorptive part of the hadron loop caused by
the final-state interactions. The dispersive part of the loop
is difficult to calculate
because of the ultraviolet divergence.  The elastic and inelastic rescattering
caused by vector trajectory( $\rho, \omega, {\rm and}\; K^*$) exchange
may give additional contributions, and there are many 
multiparticle intermiadiate states which cannot be neglected \cite{Delepine}.
Beasuse of existence of these uncertainties,
the simple Fleischer-Mannel relation $R\geq sin^{2}\gamma$ is modified
greatly by the final-state interactions. We think it is difficult to get
reasonable information about weak angle $\gamma$ in $B\to \pi K$ processes. 
The $5\%-10\%$ direct CP asymmetry can be generated by final-state 
interactions. Moreover, as discussed above, we only consider the absorptive
part of the hadron loop in this work, there are still many 
uncertaities of the theoretical predictions on the constraint of $\gamma$
and $A_{CP}^{dir}$. At present stage, there is no reliable
renormalization scheme for obtaining correct dispersive part of the hadron
loop, and it is a well-known fact for evaluating the loops in the chiral
Lagrangian theories. We will try some phenomenological ways to carry out
the renormalization elsewhere. \cite{Du}

\section*{Acknoledgement}

One of the authors (Z.T.Wei) would like to thank H.Y.Jin for 
his warm help. Li also expresses his gratitude to Dr. X.G. He for his help
and enlightening discussions.
This work is supported in part by National Natural
Science Foundation of China and the Grant of State 
Commission of Science and Technology of China.

\newpage


Table 1. The amplitude and branching ratios of the direct $B
\rightarrow \pi K$ decays. where "Tree" means only the tree 
diagram contribution; "Penguin" means only the Penguin diagram 
contribution. "Tree+Penguin" means tree plus Penguin 
diagram contributions.  

\begin{center}
\vspace{0.5cm}

\begin{tabular}{|c|c|c|c|c|c|c|} \hline \hline

 Decay Mode & $N_c$ & \multicolumn{3}{|c|} { $A^{dir}(10^{-9})$ }
  & \multicolumn{2}{|c|}  { $Br$ } \\\cline{3-7} 
 & & Tree & Penguin & Tree+Penguin & Tree & Tree+Penguin \\\hline

 
 $B_u^- \rightarrow \pi^- \bar{K^0} $ & 
 $N_c=\infty$ &
  0  &   $6.53+i36.1$   & $6.53+i36.1$  & 
  0  &   $1.18\times 10^{-5}$  \\\hline

 $\bar{B^0} \rightarrow \pi^+ K^- $ & 
 $N_c=\infty$ &
  -7.88+i2.78  &   $6.51+i34.8$   & $-1.36+i37.6$  & 
  $5.93\times 10^{-7}$  &   $1.20\times 10^{-5}$  \\ \hline \hline


 $B_u^- \rightarrow \pi^- \bar{K^0} $ & 
 $N_c=3$ &
  0  &   $5.78+i31.2$   & $5.78+i31.2$  & 
  0  &   $8.79\times 10^{-6}$  \\\hline

 $\bar{B^0} \rightarrow \pi^+ K^- $ & 
 $N_c=3$ &
  -7.16+i2.53  &   $5.82+i32.0$   & $-1.34+i34.5$  & 
  $4.9\times 10^{-7}$  &   $1.01\times 10^{-6}$  \\ \hline \hline


 $B_u^- \rightarrow \pi^- \bar{K^0} $ & 
 $N_c=2$ &
  0  &   $4.87+i27.8$   & $4.87+i27.8$  & 
  0  &   $6.94\times 10^{-6}$  \\\hline

 $\bar{B^0} \rightarrow \pi^+ K^- $ & 
 $N_c=2$ &
  -6.8+i2.4  &   $5.01+i29.7$   & $-1.79+i32.1$  & 
  $4.43\times 10^{-7}$  &   $8.81\times 10^{-6}$  \\
\hline
  \hline
\end{tabular}

\vspace{2cm}

\end{center}

\newpage
\begin{center}
Table 2. The amplitue for the final-state interactions $B\rightarrow 
\rho K^* \rightarrow \pi K$. 

\vspace{0.5cm}

\begin{tabular}{|c|c|c|c|c|c|} \hline \hline

 Decay Mode & $N_c$ & \multicolumn{3}{|c|}  { $A^{FSI}(10^{-9})$ }  
  & $A^{FSI}/A^{dir}$  \\\cline{3-5}
&  & Tree &  Penguin & Tree+Penguin & \\ \hline  


 $B_u^- \rightarrow \rho^- \bar{K^{*0}}\rightarrow \pi^- \bar{K^0}$ &
 $N_c=\infty$ &
 0  &  $9.86-i1.86$ & $9.86-i1.86$ & $-0.27e^{i89.6^0}  $ \\\hline

 $\bar{B^0} \rightarrow \rho^+ K^{*-}\rightarrow \pi^+ K^- $ &
 $N_c=\infty$ &
 $1.22+i3.47$  &  $9.35-i1.86$ & $10.6+i1.60$ & $-0.28e^{i96.5^0} $ \\\hline

 $B_u^- \rightarrow \rho^0 K^{*-}\rightarrow \pi^- \bar{K^0}$ &
 $N_c=\infty$ &
 $0.57+i1.6$  & $7.34-i2.83$  & $7.9-i1.22$ & $-0.22e^{i91.5^0}  $ \\\hline

 $\bar{B^0} \rightarrow \rho^0 \bar{K^{*0}}\rightarrow \pi^+ K^- $ & 
 $N_c=\infty$ &
 $-0.25-i0.72$  & $-5.06+i1.39$  & $-5.32+i0.67$ & $0.14e^{i80.8^0}  $ \\\hline \hline


 $B_u^- \rightarrow \rho^- \bar{K^{*0}}\rightarrow \pi^- \bar{K^0}$ &
 $N_c=3$ &
 0  &  $8.2-i1.65$ & $8.2-i1.65$  & $-0.26e^{i89.1^0} $\\\hline

 $\bar{B^0} \rightarrow \rho^+ K^{*-}\rightarrow \pi^+ K^- $ &
 $N_c=3$ &
 $1.11+i3.15$  &  $8.61-i1.67$ & $9.72+i1.49$ & $-0.28e^{i96.5^0} $ \\\hline

 $B_u^- \rightarrow \rho^0 K^{*-}\rightarrow \pi^- \bar{K^0}$ &
 $N_c=3$ &
 $0.81+i2.31$  & $6.86-i2.74$  & $7.68-i0.44$ & $-0.24e^{i97.2^0} $ \\\hline

 $\bar{B^0} \rightarrow \rho^0 \bar{K^{*0}}\rightarrow \pi^+ K^- $ & 
 $N_c=3$ &
 $0.06+i0.16$  & $-3.82+i1.11$  & $-3.77+i1.27$ & $0.12e^{i69.2^0} $\\\hline \hline


 $B_u^- \rightarrow \rho^- \bar{K^{*0}}\rightarrow \pi^- \bar{K^0}$ &
 $N_c=2$ &
 0  &  $7.1-i1.38$ & $7.1-i1.38$  & $-0.26e^{i89.0^0} $\\\hline

 $\bar{B^0} \rightarrow \rho^+ K^{*-}\rightarrow \pi^+ K^- $ &
 $N_c=2$ &
 $1.06+i2.99$  &  $8.01-i1.43$ & $9.06+i1.56$ & $-0.29e^{i96.6^0} $ \\\hline

 $B_u^- \rightarrow \rho^0 K^{*-}\rightarrow \pi^- \bar{K^0}$ &
 $N_c=2$ &
 $0.94+i2.66$  & $6.58-i2.48$  & $7.52+i0.18$ & $-0.27e^{i98.6^0} $ \\\hline

 $\bar{B^0} \rightarrow \rho^0 \bar{K^{*0}}\rightarrow \pi^+ K^- $ & 
 $N_c=2$ &
 $0.21+0.60$  & $-3.01+i0.86$  & $-2.8+i1.46$ & $0.10e^{i59.2^0} $\\
 \hline
 \hline
\end{tabular}

\end{center}

\newpage
\unitlength=0.32mm
\begin{figure}

\begin{picture}(0,0)
\put(28,200){\line(1,0){30}}
\put(62,200){\circle{8}}
\put(66,200){\line(3,2) {45}}
\put(66,200){\line(3,-2){45}}
\put(115,168){\circle{8}}
\put(115,232){\circle{8}}
\put(115,172){\line(0,1){56}}
\put(119,168){\line(1,0){38}}
\put(119,232){\line(1,0){38}}

\put(85,186.8){\vector(3,-2){2}}
\put(85,212.5){\vector(3,2){2}}

\put(10,198){$B_u^-$}
\put(72,168){$\bar{K^{*0}}$}
\put(77,220){$\rho^-$}
\put(118,195){$\pi^0$}
\put(158,228){$\pi^-$}
\put(158,163){$\bar{K^0}$}
\put(108,140){(a)}
\end{picture}

\begin{picture}(0,0)
\put(228,220){\line(1,0){30}}
\put(262,220){\circle{8}}
\put(266,220){\line(3,2) {45}}
\put(266,220){\line(3,-2){45}}
\put(315,188){\circle{8}}
\put(315,252){\circle{8}}
\put(315,192){\line(0,1){56}}
\put(319,188){\line(1,0){38}}
\put(319,252){\line(1,0){38}}
\put(285,206.8){\vector(3,-2){2}}
\put(285,232.5){\vector(3,2){2}}

\put(210,218){$\bar{B^0}$}
\put(272,188){$K^{*-}$}
\put(277,240){$\rho^+$}
\put(318,215){$\pi^0$}
\put(358,248){$\pi^+$}
\put(358,183){$K^-$}
\put(320,160){(b)}
\end{picture}

\begin{picture}(0,0)
\put(28,60){\line(1,0){30}}
\put(62,60){\circle{8}}
\put(66,60){\line(3,2) {45}}
\put(66,60){\line(3,-2){45}}
\put(115,28){\circle{8}}
\put(115,92){\circle{8}}
\put(115,32){\line(0,1){56}}
\put(119,28){\line(1,0){38}}
\put(119,92){\line(1,0){38}}
\put(85,46.8){\vector(3,-2){2}}
\put(85,72.5){\vector(3,2){2}}

\put(10,58){$B_u^-$}
\put(72,28){$K^{*-}$}
\put(77,80){$\rho^0$}
\put(118,55){$\pi^+$}
\put(158,88){$\pi^-$}
\put(158,23){$\bar{K^0}$}
\put(108,-10){(c)}
\end{picture}

\begin{picture}(0,0)
\put(228,80){\line(1,0){30}}
\put(262,80){\circle{8}}
\put(266,80){\line(3,2) {45}}
\put(266,80){\line(3,-2){45}}
\put(315,48){\circle{8}}
\put(315,112){\circle{8}}
\put(315,52){\line(0,1){56}}
\put(319,48){\line(1,0){38}}
\put(319,112){\line(1,0){38}}
\put(285,66.8){\vector(3,-2){2}}
\put(285,92.5){\vector(3,2){2}}

\put(210,78){$\bar{B^0}$}
\put(272,48){$\bar{K^{*0}}$}
\put(277,100){$\rho^0$}
\put(318,75){$\pi^-$}
\put(358,108){$\pi^+$}
\put(358,43){$K^-$}
\put(320,10){(d)}
\put(5,-30){Fig. 1.  Final-state interactoins in $B\rightarrow 
\rho K^*  \rightarrow \pi K$ due to single pion exchange.}
\put(5,-50)
{(a), (b) the neutral pion exchange case. 
(c), (d)the charged pion exchange case.}
\end{picture}


\end{figure}

\end{document}